\documentclass[%
 reprint,
 amsmath,amssymb,
 aps,
prc,
]{revtex4-2}

\usepackage{amsmath}
\usepackage{physics}
\usepackage{graphicx}% Include figure files
\usepackage{dcolumn}% Align table columns on decimal point
\usepackage{bm}% bold math
\begin{document}

%\preprint{APS/123-QED}

\title{Electromagnetic probes as signatures for a first-order QCD phase transition}

\author{Mohamad Lukman Aidid Mohd Yusoff}
\author{Norhasliza Yusof}%
\author{Hasan Abu Kassim}
\affiliation{%
 Department of Physics, 
 Faculty of Science, 
 University of Malaya,
 50603 Kuala Lumpur, Malaysia}%

\author{Jan Steinheimer}
\affiliation{GSI Helmholtzzentrum f\"ur Schwerionenforschung GmbH, Planckstr. 1, 64291 Darmstadt, Germany.}
\affiliation{
Frankfurt Institute for Advanced Studies,
Ruth-Moufang-Str. 1, 
D-60438 Frankfurt am Main, Germany}

\author{Marcus Bleicher}
\affiliation{Institut f\"{u}ur Theoretische Physik, Johann Wolfgang Goethe-Universit\"{a}t,
Max-von-Laue-Strasse 1, 60438 Frankfurt am Main, Germany}
\affiliation{Helmholtz Research Academy Hessen for FAIR (HFHF), GSI Helmholtz Center,
Campus Frankfurt, Max-von-Laue-Straße 12, 60438 Frankfurt am Main, Germany}

\author{Apiwit Kittiratpattana}
\author{Ayut Limphirat}
\author{Christoph Herold}
\email{herold@g.sut.ac.th}
\affiliation{Center of Excellence in High Energy Physics and Astrophysics, School of Physics, Suranaree University of Technology, University Avenue 111, Nakhon Ratchasima 30000, Thailand}

\date{\today}% It is always \today, today,

\begin{abstract}
We investigate dimuon production in the context of a first-order phase transition in QCD matter using a chiral fluid dynamics model. This approach incorporates non-equilibrium effects such as entropy production and reheating, which emerge during the dynamical evolution through a first-order phase transition. By comparing equilibrium and non-equilibrium scenarios across a range of beam energies ($\sqrt{s_{NN}}=2.2-6.2$~GeV), we analyze the resulting invariant mass spectra. Our results reveal a substantial enhancement of dilepton yields in the non-equilibrium scenario, particularly pronounced at lower beam energies, where reheating leads to a prolonged lifetime of the fireball and increased emission. The enhancement persists even after normalizing to pion multiplicities, indicating sensitivity beyond effects of entropy production.
\end{abstract}

\maketitle

\section{Introduction}
\label{sec:intro}

The quark-gluon plasma (QGP), a state of matter composed of deconfined quarks and gluons, is believed to have existed in the early universe between approximately $10^{-10}$ to $10^{-6}$~s after the Big Bang. Understanding the properties of this exotic phase is central to the study of the quantum chromodynamics (QCD) phase diagram and is experimentally probed through ultrarelativistic heavy-ion collisions at facilities such as the Relativistic Heavy Ion Collider (RHIC) and the Large Hadron Collider (LHC) \cite{STAR:2021iop,Friman:2011pf}.

One of the most pressing questions in heavy-ion physics concerns the nature of the phase transition between hadronic matter and the QGP. Lattice QCD simulations at vanishing baryochemical potential $\mu_B$ have established that the transition is a smooth crossover \cite{Aoki:2006we,Aoki:2006br,Aoki:2009sc,Borsanyi:2010bp}. However, at higher baryon densities, phenomenological models and functional methods, such as those based on Dyson-Schwinger equations and the functional renormalization group, suggest the existence of a first-order phase transition (FOPT) and a critical end point (CEP) in the QCD phase diagram \cite{Fischer:2014ata,Isserstedt:2019pgx,Gao:2020fbl}.

The Beam Energy Scan (BES) program at RHIC has been instrumental in searching for signatures of the CEP and the nature of the phase transition. Measurements of fluctuations and cumulants of conserved charges (such as baryon number) up to the sixth order are being used to search for critical phenomena \cite{HADES:2020wpc,STAR:2021iop,STAR:2021rls}. However, extracting these signals is challenging, as it requires high event statistics. Due to the sensitivity of these cumulants to the correlation length, it is not fully understood how much of the signal survives the non-equilibrium dynamics of the expanding medium, finite size and time effects \cite{Berdnikov:1999ph}, or hadronic rescattering, as explored by a multitude of dynamical models in recent years \cite{Athanasiou:2010kw,Stephanov:2011pb,Nahrgang:2011mg,Mukherjee:2015swa,Jiang:2015hri,Herold:2016uvv,Herold:2017day,Stephanov:2017ghc,Stephanov:2017wlw,Nahrgang:2018afz,Nahrgang:2020yxm,Du:2020bxp}. 

Dileptons offer a complementary and penetrating probe of the strongly interacting medium created in heavy-ion collisions as they are created through virtual photons throughout the evolution of the medium. Unlike hadrons, dileptons do not undergo strong final-state interactions and therefore carry undistorted information from all stages of the fireball evolution \cite{Rapp:2014hha}. Importantly, dilepton spectra have been proposed to be sensitive to the presence of a FOPT by modifying emission rates, especially in regions of low to intermediate invariant mass \cite{Seck:2020qbx}.

Non-equilibrium effects, such as the delay in phase conversion due to supercooling in a FOPT scenario, can prolong the lifetime of the system and increase dilepton production. Previous studies have demonstrated that entropy production and long-lived metastable states can leave observable imprints on dilepton spectra. These features make dileptons a promising observable to distinguish between different types of phase transitions in QCD matter.
While previous studies have also found an enhancement of dilepton production due to a non-equilibrium phase transition \cite{Seck:2020qbx,Savchuk:2022aev}, the non-equilibrium effects were never studied as explicitly as in this work.
Here, we explore dilepton production in a chiral fluid dynamics model incorporating a FOPT \cite{Herold:2018ptm,Herold:2022laa}. By comparing equilibrium and non-equilibrium evolution scenarios across a range of collision energies, we aim to identify potential signatures of a non-equilibrium FOPT in the dilepton spectra. Our goal is to understand the various contributions of typical effects of a non-equilibrium FOPT, such as entropy production, lifetime extension, and reheating.

\section{Model description}
\label{sec:model}

%%%%% SUBSECTION A
\subsection{Chiral fluid dynamics}

The spontaneous chiral symmetry breaking in vacuum as well as the restoration or chiral symmetry at large $T$ or $\mu_B$ are described by the quark-meson model with Lagrangian  
    \begin{gather}
    \label{eq:lagrangian}
    \mathcal{L}=\overline{q} \left( i\gamma ^{\mu}\partial _{\mu}-
    g\sigma \right) q
    + \frac{1}{2} \partial _{\mu}\sigma \partial ^{\mu}\sigma 
    -U(\sigma) ~,\\
    U(\sigma)=\frac{\lambda^{2}}{4} \left(\sigma ^{2}-{f_{\pi}}^{2}\right)^{2}-H\sigma ~.
    \end{gather}
Here, $q$ represents the light quark doublet $q=(u,d)$. The other parameters of this model are set to their standard values $f_\pi=93$~MeV, $m_\pi=138$~MeV and $H=f_\pi m_\pi^2$. The grand potential given in the mean-field approximation becomes
    \begin{equation}
    \label{eq:potsig}
    \Omega (T,\mu)=
    U(\sigma)+ \Omega_{q\bar{q}}~.
    \end{equation}
In this equation, the contribution of the quarks and antiquarks is given by
    \begin{equation}
    \begin{split}
        \Omega_{\bar{q}q} =& -2N_{c}N_{f}T \\
        & \times \int \frac{{\dd}^{3}p}{(2\pi)^{3}}\left\{ {\ln}\left[1+ \rm e^{\left(\frac{E-\mu}{T}\right)} \right] + {\ln}\left[1+ \rm e^{\left(\frac{E+\mu}{T}\right)} \right]\right\}~,
        \label{eq:qqbar}
    \end{split}
    \end{equation}

which resembles the negative pressure of a fermionic gas of quarks and antiquarks with quasiparticle energies $E = \sqrt{p^2 + M^2}$. The effective mass of (constituent) quarks is $M(\sigma) =  g\sigma$ with the coupling constant $g$ fixed so that $M$ in vacuum equals one third of the nucleon mass. Note that in this equation, $\mu$ is the quark chemical potential, $\mu=\mu_B/3$.

The non-equilibrium chiral fluid propagates the chiral order parameter $\sigma$ with the Langevin equation of motion \cite{Herold:2018ptm},
    \begin{equation}
     \label{eq:eom_sigma}
     \ddot \sigma+\left(\frac{D}{\tau}+\eta\right)\dot \sigma+\frac{\delta\Omega}{\delta\sigma}=\xi~.
    \end{equation}
Here, the dots refer to derivatives with respect to proper time $\tau$ and $D=1$ in the Hubble term for a one-dimensional expansion in direction of the beam axis as assumed by the Bj{\o}rken model. 

The damping coefficient $\eta$ has been calculated as \cite{Nahrgang:2011mg}
    \begin{equation}
    \label{eq:dampingcoeff}
      \eta=\frac{12 g^2}{\pi}\left[1-2n_{F}\left(\frac{m_\sigma}{2}\right)\right]\frac{1}{m_\sigma^2}\left(\frac{m_\sigma^2}{4}-M^2\right)^{3/2}~,
    \end{equation}
and the white and Gaussian noise $\xi(t)$ in Eq.~\eqref{eq:eom_sigma} satisfies the fluctuation-dissipation relation,
    \begin{equation}
    \label{eq:dissfluctsigma}
     \langle\xi(t)\xi(t')\rangle_\xi = \delta(t-t') \frac{m_{\sigma}\eta}{V} \coth\left(\frac{m_{\sigma}}{2T}\right)~.
    \end{equation}

The mass of the field $\sigma$ is given by the curvature of the thermodynamic potential at the equilibrium value $\langle\sigma\rangle$ which is equal to the second derivative of the grand potential \eqref{eq:potsig} at its minimum, 
    \begin{equation}
     \label{eq:corrl}
     m_\sigma^2=\frac{\partial^2 \Omega}{\partial\sigma^2}\bigg|_{\sigma=\langle\sigma\rangle}~,
    \end{equation}
making $m_\sigma$ effectively dependent on $T$ and $\mu$.

Assuming that the rapidity distribution of the charged particles is boost invariant within the Bjorken model, we obtain the hydrodynamic equations for the energy density $e$ and baryon number density $n$,
    \begin{gather}
    \label{eq:timeenergyden}
    \dot e = -\frac{e+p}{\tau} + \left[\frac{\delta \Omega_{\bar{q}q}}{\delta \sigma} + \left(\frac{D}{\tau}+\eta\right)\dot\sigma \right] \dot\sigma~, \\
    \dot n = -\frac{n}{\tau}~.
    \end{gather}

%%%%% SUBSECTION C
\subsection{Dimuon Production}

For the calculation of dimuon rates, we follow the explicit formulas given in \cite{Cleymans:1986na} since we are merely interested in the modification of the overall spectra from an extended lifetime and different trajectories in the QCD phase diagram and the main contribution during the evolution is expected to stem from the QGP phase. A more realistic and rigorous description using hadronic rates based on the in-medium $\rho$ and $\omega$ meson spectral functions \cite{vanHees:2007th} and the impact of a phase transition on peak positions and widths will be presented in a future work. 

\subsubsection{Rate 
in the QGP phase}

The production rate of the dimuon number $N_q$ in the QGP in phase space element $\dd^4 x \dd^4 p$ is obtained from the analytic expression
\begin{equation}
\begin{split}
    \frac{\dd N_{\rm QGP}}{\dd^{4}x \, \dd^{4}p} =& \int \frac{\dd^{3}q}{(2\pi)^{3}} \frac{\dd^{3}\bar{q}}{(2\pi)^{3}} \, v_{q\bar{q}} \, \sigma_{q\bar{q} \rightarrow \mu^+ \mu^-} \\ & \times n_{q}(q) \, n_{\bar{q}}(\bar{q}) \, \var(p - q - \bar{q})~,
\end{split}
\end{equation}
where the integral is performed over the momenta of quark, $q$, and antiquark, $\bar q$, with $v_{q\bar{q}}$ being their relative velocity, and $\sigma_{q\bar{q} \rightarrow \mu^+ \mu^-}$ denoting the electromagnetic cross section for the annihilation of a quark-antiquark pair into a dimuon,
\begin{equation}
    v_{q\bar{q}} \sigma_{q\bar{q} \rightarrow \mu^+ \mu^-} = \frac{8\pi\alpha^2}{9M} \sum_{q}{e_q^2} \left(1 + \frac{2m_{\mu}^2}{M^2}\right) \left(1 - \frac{4m_{\mu}^2}{M^2}\right)^{1/2}~.
\end{equation}

In this expression, the sum runs over the two light quark flavors with electric charge $e_q=-1/2,+2/3$, $\alpha$ is the electromagnetic coupling constant, $M$ the invariant mass of the dimuon, and $m_{\mu}$ the muon mass. The factors $n_q$ and $n_{\bar{q}}$ are the Fermi-Dirac distribution functions of the quarks and antiquarks. After evaluating the momentum space integration, one obtains 
\begin{equation}
\begin{split}
    \frac{\dd N_{\rm QGP}}{\dd^{4}x \, \dd^{4}p} =& \frac{\alpha^2}{4 \pi^2} \sum_{q}{e_q^2} \left(1 + \frac{2m_{\mu}^2}{M^2}\right) \left(1 - \frac{4m_{\mu}^2}{M^2}\right)^{1/2} \\
    &\times \exp(-E/T)K_q(p,T,\mu)~.
\end{split}
\end{equation}
Here, the function $K_q$ is defined by
\begin{gather}
\begin{split}
    K_q=&\frac{T}{p}\frac{1}{1-\exp(-E/T)} \\
    &\times \ln \frac{\left[x_2+\exp(-(E+\mu)/T)\right] \left[x_1+\exp(-\mu/T)\right]}{\left[x_1+\exp(-(E+\mu)/T)\right] \left[x_2+\exp(-\mu/T)\right]}
\end{split}~,\\
    x_1=\exp(-E_{\rm max}/T)~,~~
    x_2=\exp(-E_{\rm min}/T)~,\\
    \label{eq:emax}
    E_{\rm max}=\frac{1}{2}(E+p)~,~~
    E_{\rm min}=\frac{1}{2}(E-p)~.
\end{gather}

\begin{figure}[t]
    \centering
    \includegraphics[width=0.49\textwidth]{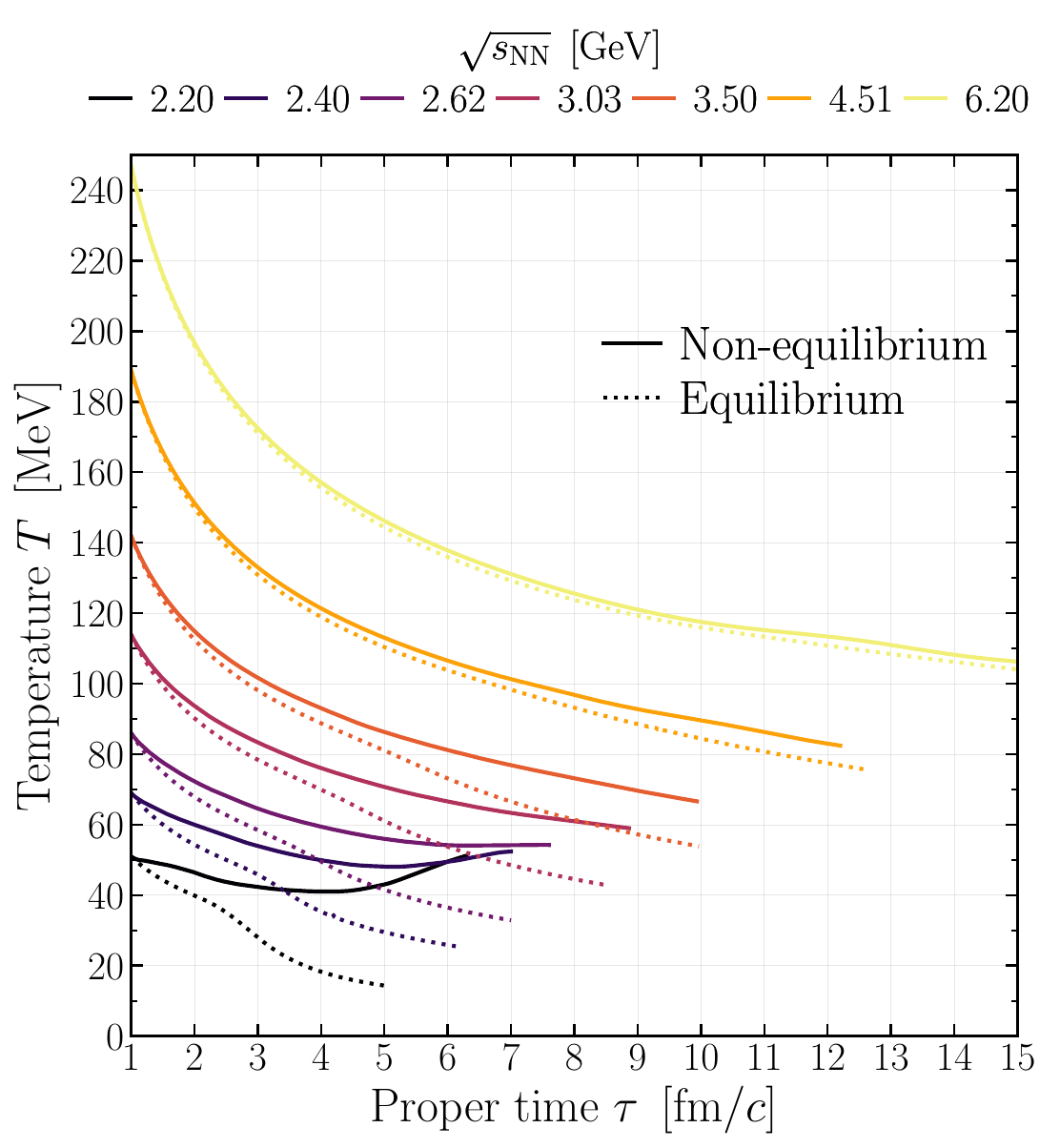}
    \caption{Temperature as function proper time for non-equilibrium (solid lines) and equilibrium (dashed lines) evolutions. The two lowest center-of-mass energies show clear reheating and lifetime extension. }
\label{fig:temp}
\end{figure}

\subsubsection{Rate in the hadronic phase}

The production rate for dimuons in the hadronic or pionic phase is treated analogously to the one in the QGP, applying an explicit analytic formula. In the Breit-Wigner approximation, the form factor of the pion is given by a sum of $\rho$-like resonances,
leading to the following approximate result for the pion form factor:
\begin{equation}
    |F_{\pi}(q^2)|^2 = \sum_{\rho} \frac{m_{\rho}^4}{(m_{\rho}^2 - q^2)^2 + m_{\rho}^2 \Gamma_{\rho}^2}~.
\end{equation}

Using the experimental values
\begin{equation}
\begin{split}
    &\quad \Gamma(\rho \rightarrow e^+ e^-) \approx 7.5 \text{ keV}, \\
    &\quad \Gamma(\rho \rightarrow \pi^+ \pi^-) \approx 60 \text{ MeV}, \\
    &\quad \Gamma(\rho \rightarrow \text{all}) \approx 260 \text{ MeV},
\end{split}
\end{equation}
we will approximate the pion form factor as:
\begin{equation}
    |F_{\pi}(q^2)|^2 \approx \frac{m_{\rho}^4}{(m_{\rho}^2 - q^2)^2 + m_{\rho}^2 \Gamma_{\rho}^2}~.
\end{equation}

Proceeding as in the of quark-antiquark annihilation in the QGP phase, one obtains the production rate in the hadronic phase using Bose-Einstein distributions for the pions,
\begin{equation}
\begin{split}
    \frac{\dd N_H}{\dd^{4}x \, \dd^{4}p} =& \frac{\alpha^2}{48 \pi^2} \left(1 + \frac{2m_{\pi}^2}{M^2}\right) \left(1 - \frac{4m_{\pi}^2}{M^2}\right)^{3/2} \\
    &\times |F_\pi(M^2)|^2 \exp(-E/T)K_H(p,T)~.
\end{split}
\end{equation}
Here, $K_H$ is adjusted from its definition for the QGP case, by changing Eq.~\eqref{eq:emax} to
\begin{gather}
    E_{\rm max}=\frac{1}{2}\left[E\left(1+\frac{m_\pi^2}{M^2}\right)+p\left(1-\frac{m_\pi^2}{M^2}\right) \right]~,\\
    E_{\rm min}=\frac{1}{2}\left[E\left(1+\frac{m_\pi^2}{M^2}\right)-p\left(1-\frac{m_\pi^2}{M^2}\right) \right]~.
\end{gather}

\section{Results}

We aim to compare the integrated dilepton rates for evolutions with a FOPT in equilibrium with the full non-equilibrium dynamics of the FOPT. 

We initialize the medium for each value of $\sqrt{s_{ \rm  NN}}$ at the corresponding point on the Rankine-Hugoniot-Taub adiabat \cite{Taub:1948zz,Thorne:1973}. 
Since the quark-meson model does not include a description of nuclear and hadronic matter, the beam energy at which the phase transition is reached is comparatively low. Here, we focus on a qualitative understanding of the processes upon reaching the non-equilibrium phase transition, for a quantitatively correct beam energy dependence one would require a much more sophisticated model for the EoS.

For the equilibrium evolution, we evolve the fluid using ideal hydrodynamics and impose that $\sigma=\langle\sigma\rangle$ at all times. In contrast to that, we use the coupled equations~\eqref{eq:eom_sigma} and \eqref{eq:timeenergyden} for the non-equilibrium evolution. In both cases the dynamics stops as soon as the freeze-out condition $\mathrm d^2 \sigma/\mathrm d \tau^2=0$ is met which is adopted from a previous work \cite{Bumnedpan:2022lma}. Fig.~\ref{fig:temp} shows for each center-of-mass energy the evolution of the temperature vs. proper time for non-equilibrium and equilibrium evolutions as solid and dashed lines, respectively. We notice two striking features: First, while the temperature of the non-equilibrium curves is generally above the equilibrium ones, a notable reheating effect occurs for the lowest two energies, i.e., $2.2$ and $2.4$~GeV, due to the sudden decay of the metastable phase resulting in a significant transfer of energy from the chiral field to the fluid. Second, for the same two energies, we find an enhanced lifetime, which is a well-known feature of a FOPT in both in equilibrium and in non-equilibrium, the latter case extending the lifetime even further. The relative differences here amount to around $15-30$\% for the nonequilibrium compared to the equilibrium case. For the other energies, the lifetimes of both scenarios are relatively similar.

\begin{figure}[t]
    \centering
    \includegraphics[width=0.49\textwidth]{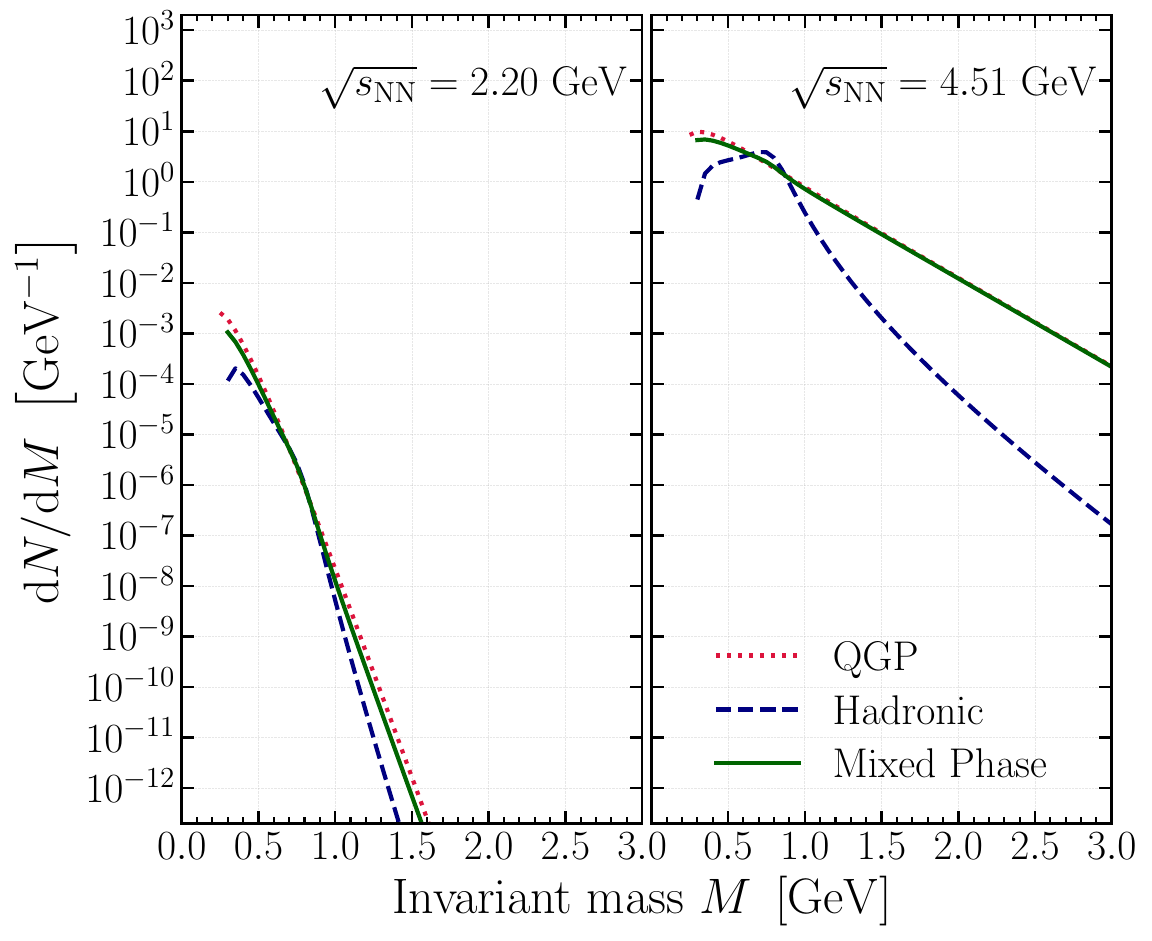}
    \caption{Comparison of QGP, hadronic, and mixed production rate of dileptons for $\sqrt{s}=2.2$ (left) and $4.51$~GeV (right).}\
\label{fig:dndm_comp}
\end{figure}

\begin{figure*}[t]
    \centering
    \includegraphics[width=0.49\textwidth]{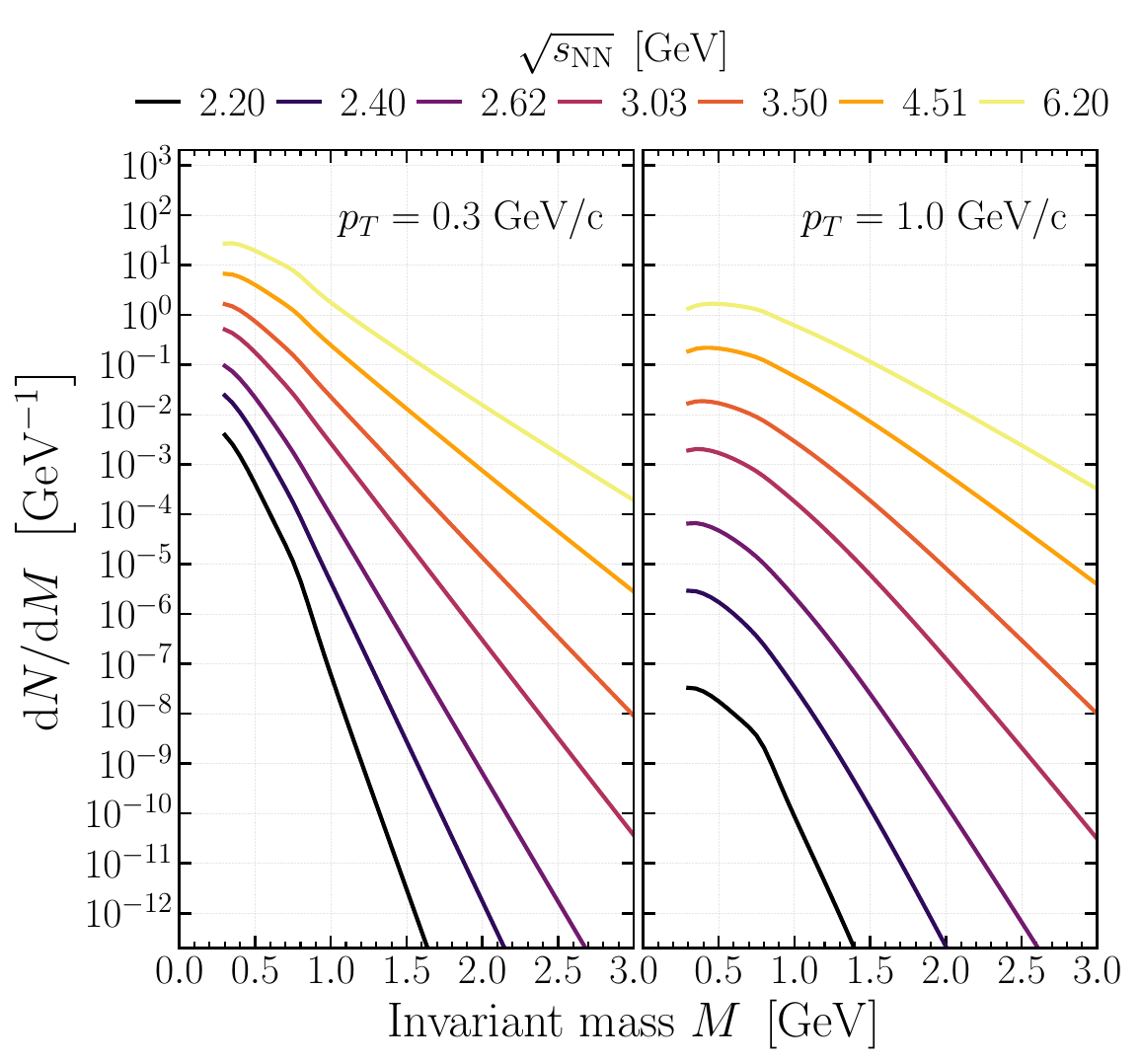}
    \includegraphics[width=0.49\textwidth]{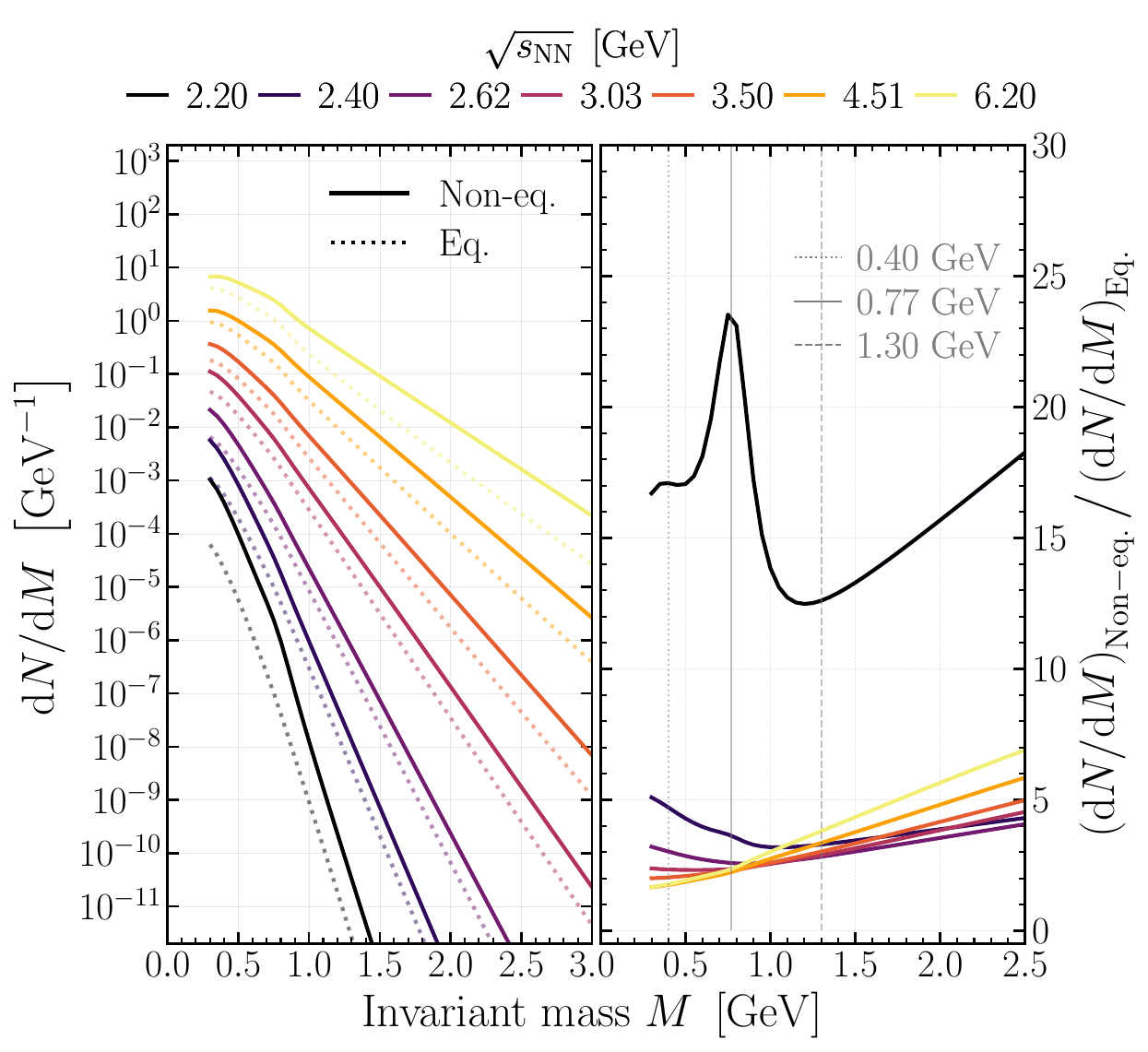}
\caption{(left) Dilepton production rate for fixed transverse momenta of $p_T=0.3$, and $1.0$~GeV for the non-equilibrium calculation. (right) Comparison of Dilepton production rates for non-equilibrium (solid lines) and equilibrium (dashed lines) scenarios with a FOPT, relative enhancement of the non-equilibrium to equilibrium scenario.}
\label{fig:dndm_pt}
\end{figure*}

We now turn to our central goal of calculating the dimuon rates in the presence of a FOPT for various energies. To describe the dynamical transition from the QGP to the hadronic phase, we use the following description for the production rate:
\begin{equation}
    \label{eq:mixed}
    \frac{\mathrm d N}{\mathrm d M}=\left[1-\frac{\sigma}{f_\pi}\right]\left(\frac{\mathrm d N}{\mathrm d M}\right)_{\rm QGP} + \frac{\sigma}{f_\pi}\left(\frac{\mathrm d N}{\mathrm d M}\right)_{\rm HG}~,
\end{equation}
which effectively serves to suppress the hadron gas rate in the QGP phase, where $\sigma\approx 0$, and the QGP rate in the hadronic phase, where $\sigma\approx f_\pi$. This mixed rate is then integrated over the evolution until the freeze-out time. 

We compare the QGP, hadronic, and mixed rate according to Eq.~\eqref{eq:mixed} in Fig.~\ref{fig:dndm_comp}, for the two energies of $\sqrt{s_{NN}}=2.2$ and $4.51$~GeV and a non-equilibrium evolution. We see that for both energies, the main component of the mixed evolution stems from the QGP part, the impact of the hadronic phase, which shows the implemented peak of the $\rho$-meson, becomes stronger at the lower energy and is practically negligible for higher ones. 

We show the dilepton production rates in non-equilibrium for all center-of-mass energies at fixed transverse momenta $p_T$ of $0.3$ and $1.0$~GeV in Fig.~\ref{fig:dndm_pt} on the left. On the right, we compare the $p_T$-integrated dilepton rates for non-equilibrium and equilibrium transitions using solid and dashed lines, respectively, for all investigated energies. We immediately see that the rates in non-equilibrium are higher than in equilibrium over the whole invariant mass range. The plot on the right quantifies the relative enhancement; we see that it generally increases with higher $M$, for the lowest energies also towards lower values of $M$. Most strikingly is the effect at $2.2$~GeV, where the rates at the non-equilibrium phase transition are roughly $15-20$ times stronger compared to the equilibrium case. In this case, we have already observed a sizable extension of the medium lifetime, thus adding more weight to contributions from the hadronic phase, as visible in the clear peak around $M$ close to the rho-meson mass for an energy of $2.2$~GeV. Still, it remains to be investigated how such a strong enhancement over all masses can occur when the lifetime is extended by at most $30$\%. For $2.4$~GeV, we observe a smaller and much less distinct structure around the rho-meson mass. 

Before continuing, let us emphasize and distinguish three effects that potentially impact the production of virtual photons during the evolution with a FOPT in non-equilibrium compared to an equilibrium scenario:
\begin{itemize}
\item the production of entropy as observed and investigated in \cite{Bumnedpan:2022lma},
\item the extended lifetime which effectively extends the time for dilepton production,
\item the reheating effect and increase in temperature, also dominant at the lowest two energies. 
\end{itemize}

\begin{figure*}[t]
    \centering
    \includegraphics[width=0.49\textwidth]{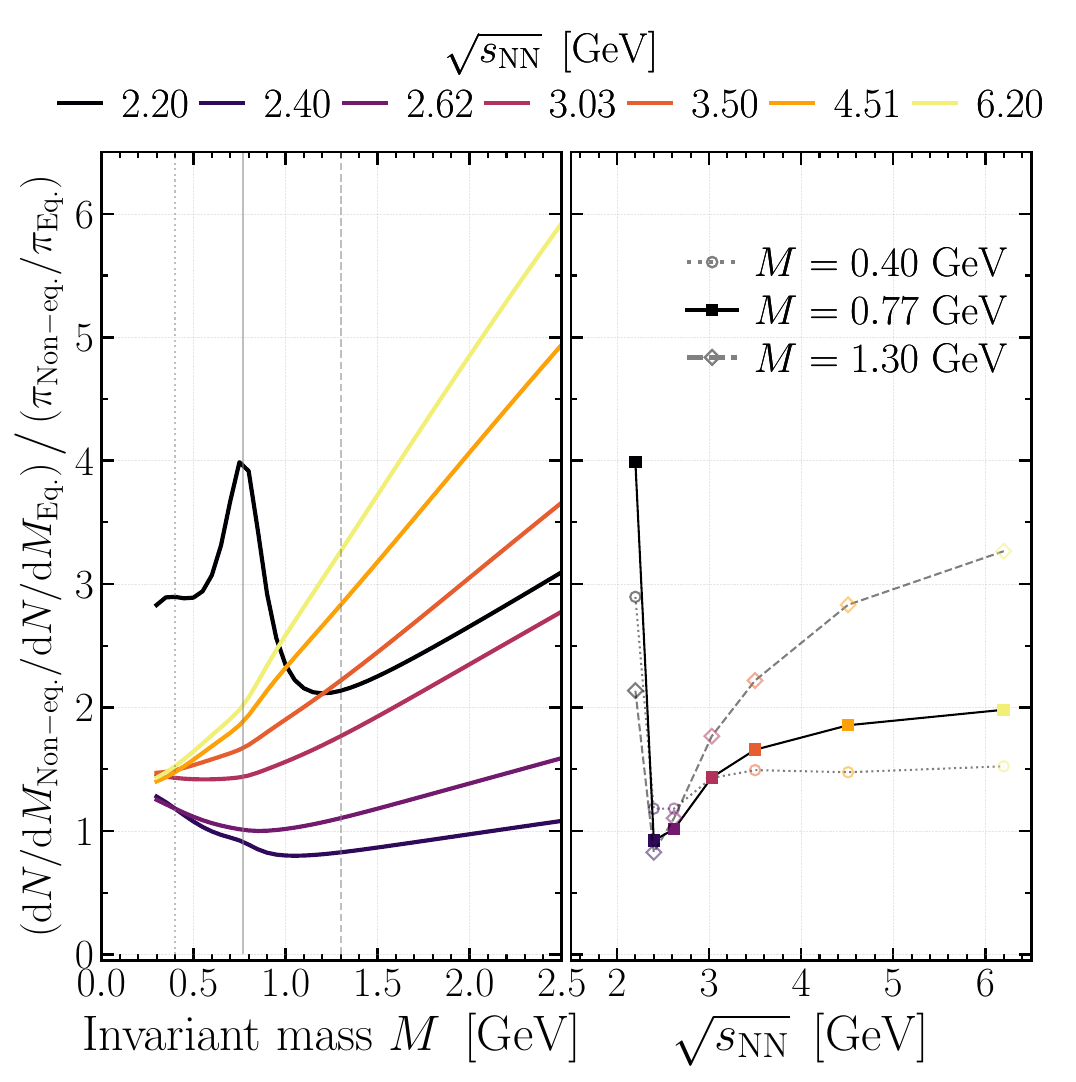}
    \includegraphics[width=0.49\textwidth]{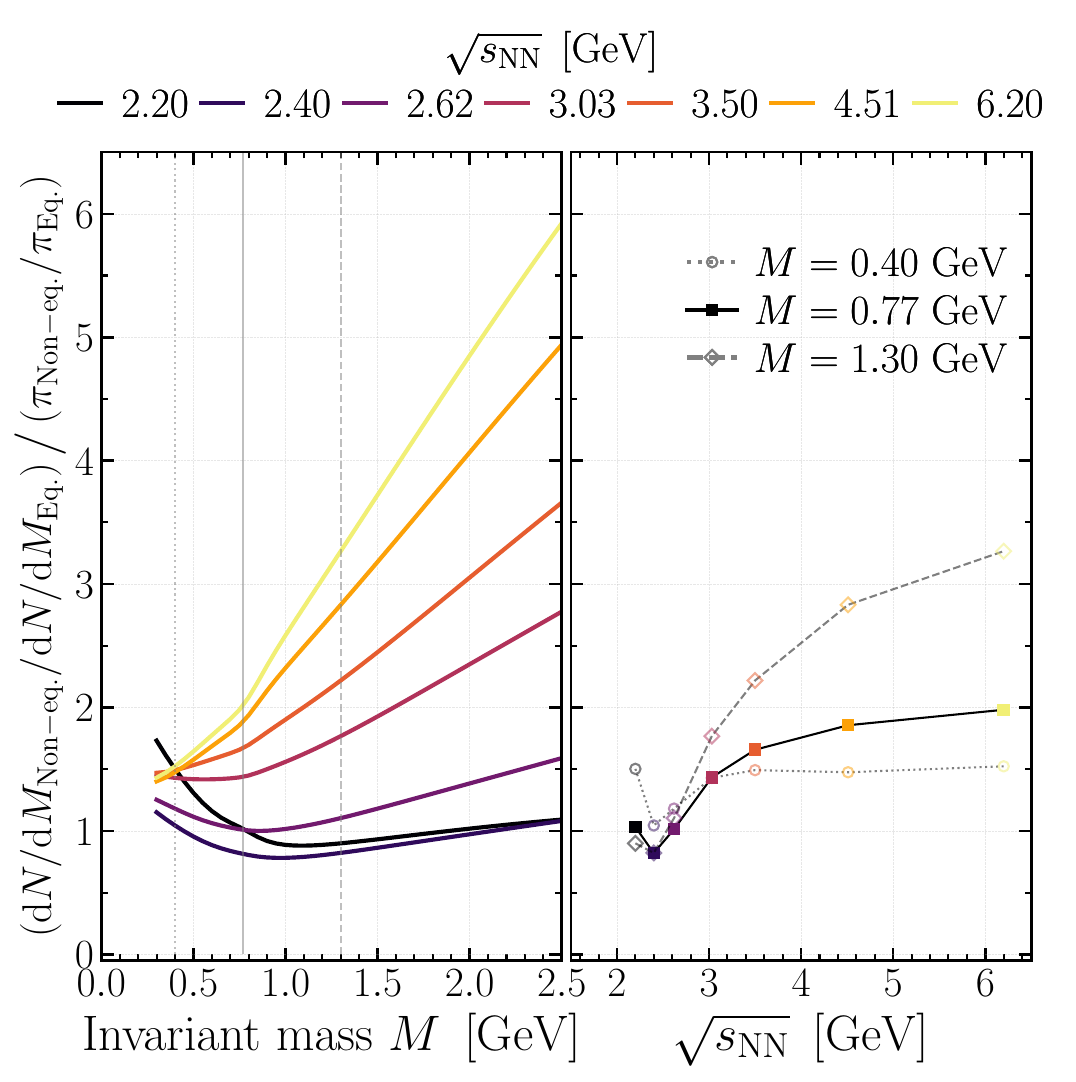}    
    \caption{Enhancement of dileptons relative to enhancement of pion production for the full evolution (left) and the scenario with equal lifetimes (right). The right hand side of each plot shows the enhancement at fixed values of $M$ as function of the energy $\sqrt{s_{\rm NN}}$.}
\label{fig:double_ratio}
\end{figure*}

We aim to understand the contribution of each of these and start by disentangling the entropy production by studying the production rates relative to the pion multiplicities. As seen in \cite{Bumnedpan:2022lma}, the ratio of entropy in non-equilibrium to equilibrium roughly resembles the ratio of pion number in non-equilibrium to equilibrium evolutions, allowing us to use
\begin{equation}
    \frac{\mathrm d N/\mathrm d M}{S}\approx \frac{\mathrm d N/\mathrm d M}{\pi}~.
\end{equation}
Fig.~\ref{fig:double_ratio} (left) shows the corresponding double ratio non-equilibrium to equilibrium as a function of the invariant mass and also for three fixed values of $M$ as function of $\sqrt{s_{\rm NN}}$. We see that the contribution of the hadronic phase persists for $\sqrt{s_{\rm NN}}=2.2$~GeV, now, however, much lower and within the same range as the ratios of all other energies. For $M>1.0$~GeV, it is surpassed by the ratios at higher center-of-mass energies of $3.5$~GeV and above. On the right hand side of the same figure, we investigate the case of equal freeze-out times for $\sqrt{s_{\rm NN}}=2.2$ and $2.4$~GeV, where we end both non-equilibrium and equilibrium evolutions at the earlier of the two freeze-out times, $\tau=5$ and $6$~fm, respectively, see Fig.~\ref{fig:temp}. Through this, we are able to mostly remove the effect of the extended lifetime in a non-equilibrium FOPT. As a result, the peak from the hadronic phase at the lowest energy has now completely disappeared, although the increase toward lower values of $M$ is still stronger than for other energies. 

The fact that we observe the strongest enhancement at such a very low energy poses challenges for its experimental detection. The initial states of our model are always created within the chirally restored phase, we know, however, from data of the flow coefficients at STAR, that at $3$~GeV, the partonic collectivity vanishes and a baryonic equation of state is favored \cite{STAR:2021yiu}. Consequently, such a strong signal would be practically unobtainable. 

\section{Conclusions}

We have explored the impact of a first-order QCD phase transition on dilepton emission using a non-equilibrium chiral fluid dynamics framework. Our model captures critical dynamical effects such as delayed phase conversion, enhanced entropy production, and - at the lowest energy - reheating, which are characteristic of a strong FOPT.

By analyzing the invariant mass spectra of dileptons across a range of low to intermediate beam energies, we have demonstrated that a full non-equilibrium FOPT leads to a significant enhancement of dilepton yields compared to an equilibrium or crossover-like scenario. This enhancement is especially pronounced at lower collision energies, at $2.2$~GeV. Here, an overall enhancement by a factor of $3-4$ compared to all other energies over the whole invariant mass range can be attributed to the entropy production in the non-equilibrium evolution through the FOPT. For lower values of $M<1.0$~GeV, the reheating and extended lifetime of the medium play a significant role, leading to enhanced production of dimuons in the hadronic phase.

\section*{Acknowledgments}

M.L.A.M.Y. and  H.A.K  acknowledge support from Universiti Malaya under Grant No. GPF044B-2018. This research has received funding support from the NSRF via the Program Management Unit for Human Resources \& Institutional Development, Research and Innovation [grant number B16F640076]. C. H. and A. L. acknowledge the support of (i) Suranaree University of Technology (ii) Thailand Science Research and Innovation (TSRI), and (iii) NSRF, project no. 195242. This work was supported by a PPP program of the DAAD.

\bibliography{mybib}

\end{document}